\documentclass[twocolumn,superscriptaddress,showpacs,amsfonts,amsmath,amssymb,prl]{revtex4} %
\usepackage{graphicx}

\begin{document}

\title{Localized states influence spin transport in epitaxial graphene}

\author{T.~Maassen}
	\email{t.maassen@rug.nl} 
\affiliation{Physics of Nanodevices, Zernike Institute for Advanced Materials, University of Groningen, Nijenborgh 4, 9747 AG Groningen, The Netherlands}
\author{J.~J.~van~den~Berg}
\affiliation{Physics of Nanodevices, Zernike Institute for Advanced Materials, University of Groningen, Nijenborgh 4, 9747 AG Groningen, The Netherlands}
\author{E.~H.~Huisman}
\affiliation{Physics of Nanodevices, Zernike Institute for Advanced Materials, University of Groningen, Nijenborgh 4, 9747 AG Groningen, The Netherlands}
\author{H. Dijkstra}
\affiliation{Physics of Nanodevices, Zernike Institute for Advanced Materials, University of Groningen, Nijenborgh 4, 9747 AG Groningen, The Netherlands}
\author{F.~Fromm}
\affiliation{Lehrstuhl f\"{u}r Technische Physik, Universit\"{a}t Erlangen-N\"{u}rnberg, Erwin-Rommel-Strasse 1, 91058 Erlangen, Germany}
\author{T.~Seyller}
	\affiliation{Lehrstuhl f\"{u}r Technische Physik, Universit\"{a}t Erlangen-N\"{u}rnberg, Erwin-Rommel-Strasse 1, 91058 Erlangen, Germany}
\author{B.~J.~van~Wees}
  \affiliation{Physics of Nanodevices, Zernike Institute for Advanced Materials, University of Groningen, Nijenborgh 4, 9747 AG Groningen, The Netherlands}%

\date{15 August 2012}

\begin{abstract}
We developed a spin transport model for a diffusive channel with coupled localized states that result in an effective increase of spin precession frequencies and a reduction of spin relaxation times in the system. We apply this model to Hanle spin precession measurements obtained on monolayer epitaxial graphene on SiC(0001) (MLEG). Combined with newly performed measurements on quasi-free-standing monolayer epitaxial graphene on SiC(0001) our analysis shows that the different values for the diffusion coefficient measured in charge and spin transport measurements in MLEG and the high values for the spin relaxation time can be explained by the influence of localized states arising from the buffer layer at the interface between the graphene and the SiC surface. 
\end{abstract}

\maketitle

\begin{figure}
\includegraphics[width=\columnwidth]{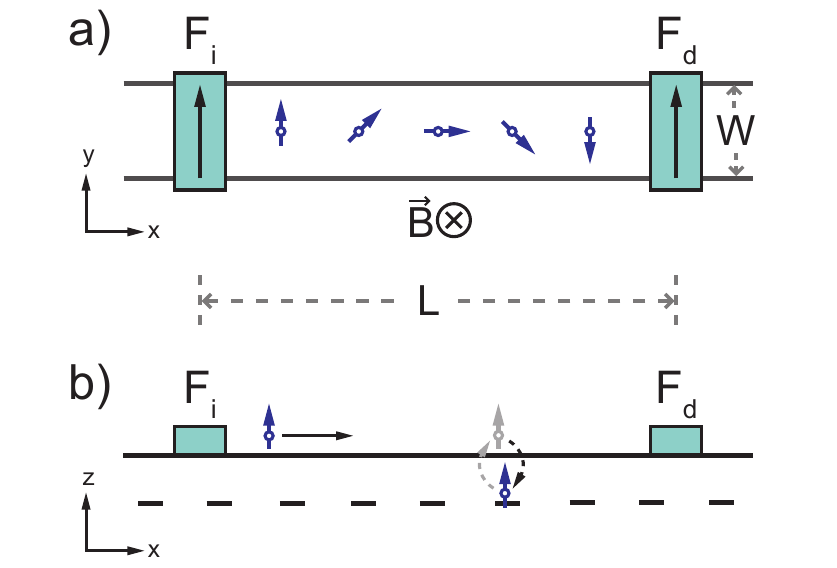} 
\caption{\label{fig:Fig1}(Color online) (a) Sketch of the Hanle precession geometry with a diffusive channel of width $W$ connected to the ferromagnetic spin injector ($F_i$) and detector ($F_d$) on distance $L$. The out-of-plane magnetic field $\vec{B}$ causes the in-plane injected spins to precess while diffusing through the channel. (b) Extension of the Hanle precession geometry with localized states that are coupled to the diffusive channel. The spins can hop into these states and back into the channel while the states are not coupled with each other.}
\end{figure}

The spin dynamics in the diffusive transport regime are in general described by the Bloch equation for the spin chemical potential $\vec{\mu_S}$ that describes the three dimensional spin accumulation \cite{APS57_Fabian2007}:
\begin{equation}
\frac{d\vec{\mu_S}}{d t} = D\mathbf\nabla^2 \vec{\mu_S} - \frac{\vec{\mu_S}}{\tau_S} + \vec{\omega_L}\times\vec{\mu_S}
\label{eq:Bloch}
\end{equation}%
with the diffusion coefficient $D$, the spin relaxation time $\tau_S$ and the Larmor frequency $\vec{\omega_L}=g\mu_B/\hbar \ \vec{B}$, that describes the spin precession in a perpendicular magnetic field $\vec{B}$ with the gyromagnetic factor $g$ (g-factor, $g=2$ for free electrons) and the Bohr magneton $\mu_B$. Experimentally, spin transport is commonly examined by Hanle spin precession measurements (Fig.~\ref{fig:Fig1}a) that are fitted with the solutions of the Bloch equation (\ref{eq:Bloch}). Those fits result in $D$, $\tau_S$ and the spin relaxation length $\lambda_S=\sqrt{D \tau_S}$. However, the fits are invariant under the transformation $D \rightarrow c \tilde{D}$, $\tau_S \rightarrow \tilde{\tau_S}/c$, $g \rightarrow c \tilde{g}$ leaving the scaling factor $c$ undefined. To unambiguously define the parameters, $D$ can be independently determined using the diffusion coefficient from charge transport measurements $D_C$ and the Einstein relation $D_C=(R_{sq} e^2 \nu(E_F))^{-1}$ \cite{foot_Weber}. Here $R_{sq}$ is the square resistance, $e$ the electron charge and $\nu(E_F)$ the density of states (DOS) of the diffusive channel at the Fermi energy.\\
Spin transport in graphene has been extensively studied in recent years \cite{N448_Tombros2007, PRB80_Popinciuc2009, PRB80_Jozsa2009, PRL105_Han2010, NL11_Avsar2011, PRL107_Han2011, PRL107_Yang2011, PRB84_Jo2011, NL12_Maassen2012, NL0_Guimaraes2012, JoVSTB30_Abel2012, c_McCreary2012, unpublished_Wojtaszek, NPaoP_Dlubak2012}. Due to weak spin-orbit coupling $g=2$ is commonly assumed to fit Hanle precession data (and define $c$) \cite{N448_Tombros2007, PRB80_Popinciuc2009, PRB80_Jozsa2009, PRL105_Han2010, NL11_Avsar2011, PRL107_Han2011, PRL107_Yang2011, PRB84_Jo2011, NL12_Maassen2012, NL0_Guimaraes2012, JoVSTB30_Abel2012, c_McCreary2012, unpublished_Wojtaszek}. This was justified for exfoliated single layer graphene (eSLG) as it was shown that $D \approx D_C$ \cite{PRB80_Jozsa2009}. On the contrary, recent results on monolayer epitaxial graphene on SiC(0001) (MLEG) \cite{PRB78_Virojanadara2008, NM8_Emtsev2009} show $D \ll D_C$ along with very high values for $\tau_S$ \cite{NL12_Maassen2012, foot_reproduced}.\\%
In this letter we introduce a model that can explain the apparent difference between $D$ and $D_C$ by the increase of the effective g-factor caused by localized states coupled to the spin transport channel. Then we discuss how this model reinterprets the results on MLEG from Ref.~\citenum{NL12_Maassen2012} and finally, we compare the results on MLEG to Hanle precession measurements on quasi-free-standing MLEG on SiC(0001) (QFMLG) \cite{unpublished_vdBerg}. In this material the graphene-like, electrical neutral buffer layer, that is in conventional MLEG located between graphene and the SiC substrate, is absent \cite{PRL103_Riedl2009, APL99_Speck2011}. The presented analysis points to the buffer layer as the origin of the localized states and by comparing the results from MLEG and eSLG we can estimate the spin properties of those states.\\%
To examine the spin transport properties of graphene, usually the non-local measurement geometry is used, consisting of a two dimensional channel with ferromagnetic electrodes that inject and detect electron spins in the graphene plane \cite{foot_tilt, N448_Tombros2007} (Fig.~\ref{fig:Fig1}a). To introduce our model we extend the description of the spin dynamics in this system with localized states in close proximity to the channel (Fig.~\ref{fig:Fig1}b). The states are electrically coupled to the channel and we assume that they are not coupled with each other. Therefore spins can hop into these states and back into the channel but not directly from one localized state into another. \\%
The spin accumulation in the localized states is represented by $\vec{\mu_S^\ast}$ and its dynamics can be described by a Bloch equation similar to $(\ref{eq:Bloch})$ that does not include a diffusive term but a term for the coupling to the transport channel.\\
\begin{equation}
\frac{d\vec{\mu_S^\ast}}{d t} = - \frac{\vec{\mu_S^\ast}}{\tau_S^\ast} + \vec{\omega_L^\ast}\times\vec{\mu_S^\ast} -\Gamma (\vec{\mu_S^\ast}-\vec{\mu_S})
\label{eq:Bloch_sI}
\end{equation}
In this equation the Lamor precession frequency in the localized states $\vec{\omega_L^\ast}=\omega_L^\ast \hat{z}\equiv \alpha \omega_L \hat{z}$ is introduced, which can be different from $\vec{\omega_L}$ due to a possibly different g-factor $g^\ast \equiv \alpha g$. $\tau_S^\ast \equiv \beta \tau_S$ is the spin relaxation time of the localized states and the term $-\Gamma(\vec{\mu_S^\ast}-\vec{\mu_S})$ describes the flow of spins from the localized states to the transport channel and vice versa with the coupling rate $\Gamma=(R e \nu_{LS})^{-1}$, where $1/R$ is the conductance per unit area between the localized states and the graphene channel and $\nu_{LS}$ the density of localized states \cite{foot_noSpinflip, foot_SM}.\\
To describe the spin dynamics in the transport channel we also have to add on the right side of the Bloch equation~(\ref{eq:Bloch}) a coupling term in the same way as in equation (\ref{eq:Bloch_sI}). Therefore we get 
\begin{equation}
\frac{d\vec{\mu_S}}{d t} = D\mathbf\nabla^2 \vec{\mu_S} - \frac{\vec{\mu_S}}{\tau_S} + \vec{\omega_L}\times\vec{\mu_S}-\eta \Gamma (\vec{\mu_S}-\vec{\mu_S^\ast}).
\label{eq:Bloch_s}
\end{equation}%
Here we introduce the factor $\eta \equiv \nu_{LS}/\nu_{gr}$ that accounts for the different DOS in graphene $\nu_{gr}$ compared to the localized states.\\
The two coupled equations (\ref{eq:Bloch_sI}) and (\ref{eq:Bloch_s}) can be reduced to \textit{one} effective Bloch equation. For this purpose we consider the system to be in a steady state with $\frac{d\vec{\mu_S^\ast}}{d t} = 0$, so that equation (\ref{eq:Bloch_sI}) can be rewritten as $\vec{\mu_S^\ast}=\underline{a} \cdot \vec{\mu_S}$ with
\begin{equation}
\begin{split}
\underline{a}= \ & \frac{\tau_S^\ast \Gamma}{(\tau_S^\ast \Gamma +1)^2+(\tau_S^\ast \omega_L^\ast)^2} \\%
& \times \left(\begin{array}{ccc} \tau_S^\ast \Gamma +1 & - \tau_S^\ast \omega_L^\ast & 0 \\
												  \tau_S^\ast \omega_L^\ast & \tau_S^\ast \Gamma +1 & 0 \\
												 0 & 0 & \tau_S^\ast \Gamma +1 + \frac{(\tau_S^\ast \omega_L^\ast)^2}{\tau_S^\ast \Gamma +1} \end{array}\right).
\end{split}												 
\label{eq:a}
\end{equation}
As the spin accumulation is purely perpendicular to the magnetic field in the Hanle geometry \cite{foot_tilt} ($\vec{\omega_L}\ ||\ \vec{\omega_L^\ast}\ ||\ \vec{B}\ ||\ \hat{z}\ \bot\ \vec{\mu_S}$) we get the effective Bloch equation
\begin{equation}
0 = D\mathbf\nabla^2 \vec{\mu_S} - \frac{\vec{\mu_S}}{\tau^{eff}_S} + \vec{\omega^{eff}_L}\times\vec{\mu_S}.
\label{eq:Bloch_eff}
\end{equation}
Here we introduce the effective spin relaxation time $\tau^{eff}_S$ and the effective precession frequency of the system $\vec{\omega^{eff}_L}=\omega^{eff}_L \hat{z}$ defined by 
\begin{eqnarray}
\frac{1}{\tau^{eff}_S} &=& \frac{1}{\tau_S} + \eta \Gamma \frac{1+\tau_S^\ast \Gamma + (\tau_S^\ast \omega_L^\ast)^2}{(1+\tau_S^\ast \Gamma)^2+(\tau_S^\ast \omega_L^\ast)^2}\label{eq:tau_eff}\\
\mathrm{and} \quad \omega^{eff}_L &=& \omega_L + \eta \Gamma^2 \frac{(\tau_S^\ast)^2 \omega_L^\ast}{(1+\tau_S^\ast \Gamma)^2 + (\tau_S^\ast \omega_L^\ast)^2}.\label{eq:omega_eff}
\end{eqnarray}
So the effective spin dynamics in the transport channel including the localized states can be described by one single effective Bloch equation with an effective spin relaxation time and precession frequency.\\
\begin{figure}
\includegraphics[width=\columnwidth]{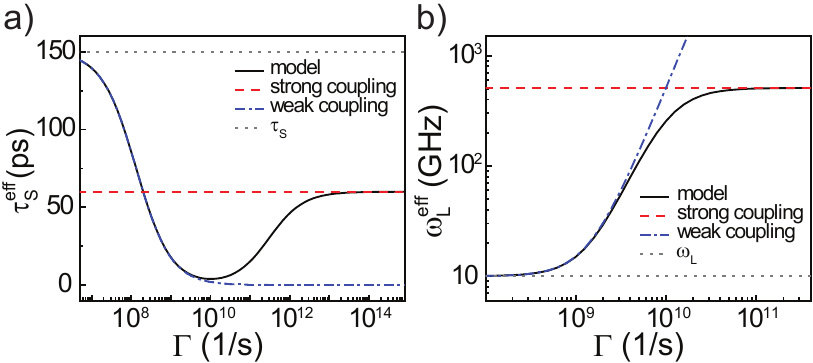} 
\caption{\label{fig:Fig2}(Color online) The effective spin relaxation time $\tau^{eff}_S$ (a) and Lamor precession frequency $\omega^{eff}_L$ (b) as a function of the coupling rate $\Gamma$ (black solid lines). The asymptotic values in the limit of strong (red, dashed line) and weak coupling (blue, dash dotted line). In the shown graphs we keep $\tau_S=150~\mathrm{ps}$ (gray dotted line in panel (a)), $\tau_S^\ast=5~\mathrm{ns}$, $\omega_L=10~\mathrm{GHz}$ (for a magnetic field of $B\approx 50~\mathrm{mT}$, gray dotted line in panel (b)) and $\eta=50$ constant. } 
\end{figure}%
The expressions for $\tau^{eff}_S$ and $\omega^{eff}_L$ are plotted in Fig.~\ref{fig:Fig2} as a function of the coupling rate $\Gamma$ (black solid lines). Note that independent from $\Gamma$ the model shows an effective decrease of the spin relaxation time (\ref{eq:tau_eff}) and an effectively increased precession frequency (\ref{eq:omega_eff}) as the coupled localized states result in extra relaxation and additional spin precession. To further analyze the effective values we can consider certain limits. $\Gamma$ is inversely proportional to the average dwell time in the localized states $\Gamma \propto 1/\tau_{dwell}$, the average time the spins stay in the localized states before hopping back into the channel. If we consider weak coupling, $\Gamma \ll 1/\tau_S^\ast$ (Fig.~\ref{fig:Fig2}, blue dash dotted lines), we have long dwell times and therefore $\tau_{dwell}\gg\tau_S^\ast$. As a consequence all the spins that hop into the localized states will relax before returning into the diffusive channel and are therefore ``lost'' for the spin transport. We get $1/\tau^{eff}_S \approx 1/\tau_S+\eta \Gamma$, while $\omega^{eff}_L = \omega_L + \mathcal{O}((\tau_S^\ast \Gamma)^2)$ stays approximately constant. \\
If we have strong coupling, $\Gamma \gg 1/\tau_S^\ast$ (Fig.~\ref{fig:Fig2}, red dashed lines), we have to distinguish between the cases where the precession frequency in the localized states is greater or lower than the coupling rate. In the case that the precession frequency is higher, $\omega_L^\ast \gg \Gamma$, we get the same result for $\tau^{eff}_S$ and $\omega^{eff}_L$ as for weak coupling. The strong precession in the localized states dephases all spins that hop into these states so that again all spins hopping into these states are lost. The influence on the effective precession frequency is therefore marginal. \\
The most interesting case is the case of strong coupling, $\Gamma \gg 1/\tau_S^\ast$, and low precession frequencies, $\omega_L^\ast \ll \Gamma$, as this corresponds to the measurements in MLEG (see below). We get: $1/\tau^{eff}_S=1/\tau_S+\eta/\tau_S^\ast$ and $\omega^{eff}_L=\omega_L + \eta \omega_L^\ast$. Both values are in this limit independent from the coupling rate $\Gamma$ (Fig.~\ref{fig:Fig2}). \\
%
%
How does this model relate to the results on spin transport in MLEG on SiC(0001) reported in Ref.~\citenum{NL12_Maassen2012}? Here an increased $\tau_S$ and a strongly reduced diffusion coefficient ($D \ll D_C$) were observed. To understand this observation in correspondence to the discussed model, let us revisit how the spin transport data is acquired. As mentioned before $g = 2$ is commonly assumed to fit Hanle precession measurements on graphene. In our system with an effective precession frequency $\omega^{eff}_L \equiv \xi \omega_L > \omega_L$ and hence $g^{eff} \equiv \xi g$ this assumption presents itself wrong. The values that are received by fitting assuming $g = 2$ are described by a modified Bloch equation that we receive by dividing (\ref{eq:Bloch_eff}) by the scaling factor $\xi$:
\begin{equation}
0 = D^{mod}\mathbf\nabla^2 \vec{\mu_S} - \frac{\vec{\mu_S}}{\tau^{mod}_S} + \vec{\omega_L}\times\vec{\mu_S}\label{eq:Bloch_meas}
\end{equation}%
with $D^{mod}=D/\xi$ and $\tau^{mod}_S=\xi \tau^{eff}_S$. The effective spin relaxation time of the system including the localized states can be obtained by either assuming $D=D_C$ for the fit or assuming $g=2$ and correcting the spin relaxation time with $\tau^{eff}_S=\tau^{mod}_S/\xi$. The enhanced value measured for $\tau_S$ in Ref.~\citenum{NL12_Maassen2012} is therefore not an intrinsic property of MLEG on SiC(0001) but is based on assuming a value for $g$ of graphene without taking the influence of the localized states into account. \\
Note that the measured spin relaxation length does not change when assuming a different g-factor as $\lambda_S^{mod}=(D^{mod} \tau^{mod}_S)^{1/2} = (D \tau^{eff}_S)^{1/2} = \lambda_S^{eff}$. Hence, the spin relaxation length in the system is reduced by the influence of the localized states.\\
\begin{figure}
\includegraphics[width=\columnwidth]{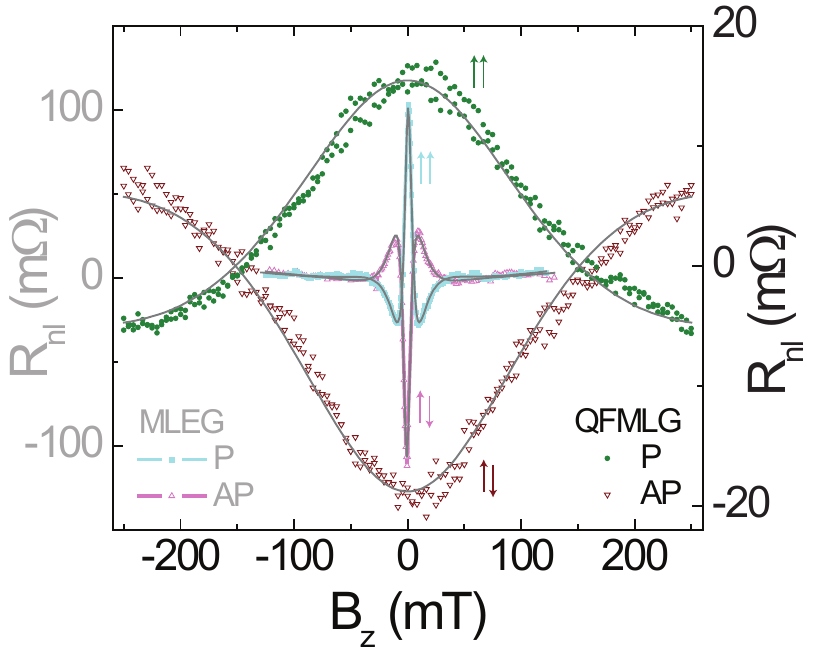} 
\caption{\label{fig:Fig3}(Color online) Hanle precession measurements performed at RT with parallel ($\uparrow\uparrow$) and antiparallel alignment ($\uparrow\downarrow$) of the inner electrodes. The measurements with the narrow curve in the middle (light colors, scale on the left axis) is taken from Ref.~\citenum{NL12_Maassen2012} and were performed on MLEG on a $L=1.2~\mathrm{\mu m}$ long and $W=0.7~\mathrm{\mu m}$ wide strip. The broader measurements enclosing the MLEG measurements (dark colors, scale on the right axis) were performed on QFMLG with $L=1.5~\mathrm{\mu m}$ and $W=1~\mathrm{\mu m}$. The fits to the solutions of the Bloch equation (\ref{eq:Bloch}) are plotted in gray. For both sets of measurements a constant background resistance was subtracted.}
\end{figure}%
To show the effect of a modified g-factor let us revisit the data in Ref.~\citenum{NL12_Maassen2012}. The narrow Hanle precession measurement in the center of Fig.~\ref{fig:Fig3} is performed on MLEG on SiC(0001) at room temperature (RT). The fit to this data (assuming $g=2$) gives $\tau_S=1.3~\mathrm{ns}$ and $D=2.4~\mathrm{cm^2/s}$, resulting in $\lambda_S=0.56~\mathrm{\mu m}$. This fit shows the increased value of $\tau_S$ compared to values obtained on eSLG \cite{NL12_Maassen2012, PRB80_Jozsa2009} but also the strongly reduced values for $D$ compared to the value of $D_C \approx 190~\mathrm{cm^2/s}$ obtained in charge transport measurements on the same sample and compared to $D \sim 200~\mathrm{cm^2/s}$ typically measured on eSLG \cite{PRB80_Jozsa2009}. \\
To find out where the predicted localized states originate from, we prepared and measured spin transport samples on quasi-free-standing MLEG (QFMLG). To obtain QFMLG only the electrical neutral buffer layer and no graphene layers are grown on SiC(0001) and the sample is then intercalated by hydrogen as described in Refs.~\citenum{PRL103_Riedl2009} and \citenum{APL99_Speck2011}. Thus the buffer layer decouples from the substrate and converts into a graphene layer while dangling bonds of the SiC surface are passivated. We are therefore left with a single graphene layer directly on the SiC(0001) surface. The experiments were performed in the same way as described for MLEG samples in Ref.~\citenum{NL12_Maassen2012}. Fig.~\ref{fig:Fig3} shows a Hanle precession curve measured on a QFMLG strip at RT next to the measurement on MLEG from Ref.~\citenum{NL12_Maassen2012}. This Hanle curve shows clearly a big difference compared to the measurement on MLEG on SiC(0001) in the same figure. The non-local resistance \cite{N448_Tombros2007, NL12_Maassen2012} changes slower with the magnetic field \cite{foot_length}, comparable to measurements performed on eSLG \cite{PRB80_Jozsa2009}. The fit (assuming $g=2$) gives $\tau_S=33.6\pm0.9~\mathrm{ps}$ and $D=75\pm2~\mathrm{cm^2/s}$ and therefore $\lambda_S=0.50\pm0.01~\mathrm{\mu m}$.\\
These values are similar to a low quality eSLG sample as $\tau_S$ is reduced by about a factor $4$ and $D$ by a factor of approximately $3$ compared to typical values obtained on eSLG \cite{PRB80_Jozsa2009}. Compared to MLEG we see an increase of D by a factor of $\sim 30$ and a decrease of $\tau_S$ by about $40$ times. And in contrast to the data obtained on MLEG we see in charge transport measurements on similar QFMLG samples a diffusion coefficient of $D_C\approx45~\mathrm{cm^2/s} \sim D$ \cite{foot_DDc}. To obtain $D_C$ we use the square resistance of $R_{sq}\approx 3.5~\mathrm{k\Omega}$ and a hole charge carrier density of $p \approx6 \times 10^{12} ~\mathrm{cm^{-2}}$ obtained in Hall measurements consistent with results from Ref.~\citenum{APL99_Speck2011}.\\
Comparing the results on the two types of graphene on SiC(0001) it is interesting to see the striking difference of the spin relaxation times and diffusion coefficients obtained in spin transport measurements but even more important that we see $D_C \approx D$ in QFMLG as we would expect for graphene \cite{PRB80_Jozsa2009}. This points to the fact that here is no effect of the localized states. Therefore, the localized states have their origin in the interface between the graphene layer and the SiC substrate as this is the only structural property that is altered between conventional MLEG and QFMLG. Hence, the states could be in the dangling bonds or in the buffer layer. The strong difference of $D$ vs $D_C$ reported in Ref.~\citenum{NL12_Maassen2012} points to a strong change in $\omega_L$. Within our model, this is the case for strong coupling of the localized states to the channel (see eq. (\ref{eq:omega_eff}) and Fig.~\ref{fig:Fig2}b). The coupling is strongest if the localized states are located in the buffer layer as this one is closest to the channel. If we assume that the coupling of the states in the buffer layer to the channel is comparable to the coupling between adjacent layers in graphite and considering $\eta \sim 50$ (see below) we get $\Gamma \sim 2  \times 10^{13} s^{-1}$ \cite{foot_SM}. For this value the strong coupling limit seems justified (see Fig.~\ref{fig:Fig2}).\\ 
Now we can evaluate the model and characterize the localized states by comparing the fitting results on MLEG on SiC(0001) from Ref.~\citenum{NL12_Maassen2012} with data obtained on other types of monolayer graphene. To compensate for different $D_C$-values obtained in charge transport measurements on QFMLG and conventional MLEG, we use data on eSLG \cite{PRB80_Jozsa2009} to compare with the fitting results on conventional MLEG \cite{NL12_Maassen2012}. In the limit of strong coupling we get based on (\ref{eq:Bloch_meas}): $\xi=1+\alpha \eta$. We assume that the g-factor in the localized states is equal to the graphene channel ($\alpha=1$). Using the typical eSLG values, $\tau_S=150~\mathrm{ps}$ and $D_C=D=200~\mathrm{cm^2/s}$, as the graphene values in the absence of localized states and the values obtained on MLEG as $D^{mod}$ and $\tau_S^{mod}$ we get $\xi \approx \eta \approx 80$ at RT. With this result and $\tau_S^{mod}/\tau_S\approx 9$ we obtain $\tau_S^\ast/\tau_S=\beta \approx 10$. Hence, at RT spins relax in the localized states with $\tau_S^\ast\approx 1.5~\mathrm{ns}$ about $10$ times slower than in the graphene channel. This enhanced value is very reasonable for a confined state in a material with low spin-orbit coupling.\\
The presence of localized states can also explain the temperature dependence of the spin transport properties in Ref.~\citenum{NL12_Maassen2012} in contrast to only negligible change for eSLG \cite{N448_Tombros2007}. By assuming the same values as before as the values of $D$ and $\tau_S$ in the absence of localized states, we get with the data for MLEG at $4~\mathrm{K}$ $\eta \approx 45$ and $\beta \approx 22$ ($\tau_S^\ast\approx 3.3~\mathrm{ns}$). These results imply that at low temperature there are less localized states accessible and those states have a longer spin relaxation time. By assuming a Boltzmann distribution we get from the change in $\eta$ an activation energy for the states of $E_a\approx15~\mathrm{meV}$.\\
Within our analysis, $\eta$ describes the ratio of the DOS in the localized states and the graphene channel. With $\eta$ up to $80$ we need a high density of localized states in our system. In MLEG we have with a electron charge carrier density of $n \approx 3 \times 10^{12}~\mathrm{cm^{-2}}$ \cite{NL12_Maassen2012} a DOS of $\nu_{gr} \approx 3 \times10^{13}~\mathrm{eV^{-1}cm^{-2}}$. With a density of carbon atoms in the graphene-like buffer layer of $3.8 \times 10^{15}~\mathrm{cm^{-2}}$ and assuming that every carbon atom contributes one localized state, we get $\eta=80$ if these states are e.g. uniformly distributed over an energy range of $\approx 1~\mathrm{eV}$. Those localized states can be the origin of the strong doping observed in MLEG on SiC(0001) \cite{APL97_Kopylov2010}.\\
The observed increase of $g$ in MLEG could in principle also be related to magnetic moments induced by the buffer layer or dangling bonds on the surface of SiC(0001) as described for hydrogenated graphene in Ref.~\citenum{c_McCreary2012}. We argue that this does not apply here since: i)~The effect in MLEG is stronger at RT than at $4~\mathrm{K}$ while the effect in hydrogenated graphene is only observed at low temperature. ii)~We do not see any effects resulting from randomized magnetic moments at low magnetic fields like the ``dip'' in the spin-valve measurements in Ref.~\citenum{c_McCreary2012}. iii)~The increase of $g$ in MLEG is much bigger than in hydrogenated graphene.\\ 
%
To summarize, we developed a spin transport model for a diffusive channel with coupled localized states that results in an increased effective g-factor and a reduced spin relaxation time for the transported spins. This model reinterprets the data from Ref.~\citenum{NL12_Maassen2012} where an enhanced spin relaxation time and a reduced spin diffusion coefficient were observed. By comparing the data from Ref.~\citenum{NL12_Maassen2012} to new measurements on QFMLG and typical values on eSLG we could identify the buffer layer as possible source for the localized states and the measurements can be related to a g-factor of $g_{eff}=(45 - 80) g$. Finally we use the model to characterize the spin properties of the localized states in the buffer layer of MLEG on SiC(0001).\\

\begin{acknowledgments}
We would like to acknowledge H.~M.~de~Roosz, B.~Wolfs, and J.~G.~Holstein for technical support and F.~L.~Bakker, I.~J.~Vera-Marun and J.~Fabian for helpful discussions. The research leading to these results has received funding from NanoNed, the Zernike Institute for Advanced Materials, the Foundation for Fundamental Research on Matter (FOM), the Deutsche Forschungsgemeinschaft and the European Union Seventh Framework Programme (FP7/2007-2013) under grant agreement ``ConceptGraphene'' Number 257829.
\end{acknowledgments}
%
%
%
%

\begin{thebibliography}{29}
\expandafter\ifx\csname natexlab\endcsname\relax\def\natexlab#1{#1}\fi
\expandafter\ifx\csname bibnamefont\endcsname\relax
  \def\bibnamefont#1{#1}\fi
\expandafter\ifx\csname bibfnamefont\endcsname\relax
  \def\bibfnamefont#1{#1}\fi
\expandafter\ifx\csname citenamefont\endcsname\relax
  \def\citenamefont#1{#1}\fi
\expandafter\ifx\csname url\endcsname\relax
  \def\url#1{\texttt{#1}}\fi
\expandafter\ifx\csname urlprefix\endcsname\relax\def\urlprefix{URL }\fi
\providecommand{\bibinfo}[2]{#2}
\providecommand{\eprint}[2][]{\url{#2}}

\bibitem[{\citenamefont{Fabian et~al.}(2007)\citenamefont{Fabian,
  Matos-Abiague, Ertler, Stano, and Zutic}}]{APS57_Fabian2007}
\bibinfo{author}{\bibfnamefont{J.}~\bibnamefont{Fabian}},
  \bibinfo{author}{\bibfnamefont{A.}~\bibnamefont{Matos-Abiague}},
  \bibinfo{author}{\bibfnamefont{C.}~\bibnamefont{Ertler}},
  \bibinfo{author}{\bibfnamefont{P.}~\bibnamefont{Stano}}, \bibnamefont{and}
  \bibinfo{author}{\bibfnamefont{I.}~\bibnamefont{Zutic}},
  \bibinfo{journal}{Acta Phys. Slov.} \textbf{\bibinfo{volume}{57}},
  \bibinfo{pages}{565} (\bibinfo{year}{2007}).

\bibitem[{foo({\natexlab{a}})}]{foot_Weber}
\bibinfo{note}{Note that for a diffusive channel generally impurity scattering
  dominates and $D=D_C$ holds, while a difference can arise due to strong
  electron-electron interactions e.g. in a two-dimensional electron gas as
  discussed by Weber et al., Nature (London) \textbf{437}, 1330 (2005)}.

\bibitem[{\citenamefont{Tombros et~al.}(2007)\citenamefont{Tombros, J\'{o}zsa,
  Popinciuc, Jonkman, and van Wees}}]{N448_Tombros2007}
\bibinfo{author}{\bibfnamefont{N.}~\bibnamefont{Tombros}},
  \bibinfo{author}{\bibfnamefont{C.}~\bibnamefont{J\'{o}zsa}},
  \bibinfo{author}{\bibfnamefont{M.}~\bibnamefont{Popinciuc}},
  \bibinfo{author}{\bibfnamefont{H.~T.} \bibnamefont{Jonkman}},
  \bibnamefont{and} \bibinfo{author}{\bibfnamefont{B.~J.} \bibnamefont{van
  Wees}}, \bibinfo{journal}{Nature (London)} \textbf{\bibinfo{volume}{448}},
  \bibinfo{pages}{571} (\bibinfo{year}{2007}).

\bibitem[{\citenamefont{Popinciuc et~al.}(2009)\citenamefont{Popinciuc,
  J\'ozsa, Zomer, Tombros, Veligura, Jonkman, and van
  Wees}}]{PRB80_Popinciuc2009}
\bibinfo{author}{\bibfnamefont{M.}~\bibnamefont{Popinciuc}},
  \bibinfo{author}{\bibfnamefont{C.}~\bibnamefont{J\'ozsa}},
  \bibinfo{author}{\bibfnamefont{P.~J.} \bibnamefont{Zomer}},
  \bibinfo{author}{\bibfnamefont{N.}~\bibnamefont{Tombros}},
  \bibinfo{author}{\bibfnamefont{A.}~\bibnamefont{Veligura}},
  \bibinfo{author}{\bibfnamefont{H.~T.} \bibnamefont{Jonkman}},
  \bibnamefont{and} \bibinfo{author}{\bibfnamefont{B.~J.} \bibnamefont{van
  Wees}}, \bibinfo{journal}{Phys. Rev. B} \textbf{\bibinfo{volume}{80}},
  \bibinfo{pages}{214427} (\bibinfo{year}{2009}).

\bibitem[{\citenamefont{J\'{o}zsa et~al.}(2009)\citenamefont{J\'{o}zsa,
  Maassen, Popinciuc, Zomer, Veligura, Jonkman, and van
  Wees}}]{PRB80_Jozsa2009}
\bibinfo{author}{\bibfnamefont{C.}~\bibnamefont{J\'{o}zsa}},
  \bibinfo{author}{\bibfnamefont{T.}~\bibnamefont{Maassen}},
  \bibinfo{author}{\bibfnamefont{M.}~\bibnamefont{Popinciuc}},
  \bibinfo{author}{\bibfnamefont{P.~J.} \bibnamefont{Zomer}},
  \bibinfo{author}{\bibfnamefont{A.}~\bibnamefont{Veligura}},
  \bibinfo{author}{\bibfnamefont{H.~T.} \bibnamefont{Jonkman}},
  \bibnamefont{and} \bibinfo{author}{\bibfnamefont{B.~J.} \bibnamefont{van
  Wees}}, \bibinfo{journal}{Phys. Rev. B} \textbf{\bibinfo{volume}{80}},
  \bibinfo{eid}{241403} (\bibinfo{year}{2009}).

\bibitem[{\citenamefont{Han et~al.}(2010)\citenamefont{Han, Pi, McCreary, Li,
  Wong, Swartz, and Kawakami}}]{PRL105_Han2010}
\bibinfo{author}{\bibfnamefont{W.}~\bibnamefont{Han}},
  \bibinfo{author}{\bibfnamefont{K.}~\bibnamefont{Pi}},
  \bibinfo{author}{\bibfnamefont{K.~M.} \bibnamefont{McCreary}},
  \bibinfo{author}{\bibfnamefont{Y.}~\bibnamefont{Li}},
  \bibinfo{author}{\bibfnamefont{J.~J.~I.} \bibnamefont{Wong}},
  \bibinfo{author}{\bibfnamefont{A.~G.}~\bibnamefont{Swartz}}, \bibnamefont{and}
  \bibinfo{author}{\bibfnamefont{R.~K.} \bibnamefont{Kawakami}},
  \bibinfo{journal}{Phys. Rev. Lett.} \textbf{\bibinfo{volume}{105}},
  \bibinfo{pages}{167202} (\bibinfo{year}{2010}).

\bibitem[{\citenamefont{Avsar et~al.}(2011)\citenamefont{Avsar, Yang, Bae,
  Balakrishnan, Volmer, Jaiswal, Yi, Ali, G\"{u}ntherodt, Hong
  et~al.}}]{NL11_Avsar2011}
\bibinfo{author}{\bibfnamefont{A.}~\bibnamefont{Avsar}},
  \bibinfo{author}{\bibfnamefont{T.-Y.} \bibnamefont{Yang}},
  \bibinfo{author}{\bibfnamefont{S.}~\bibnamefont{Bae}},
  \bibinfo{author}{\bibfnamefont{J.}~\bibnamefont{Balakrishnan}},
  \bibinfo{author}{\bibfnamefont{F.}~\bibnamefont{Volmer}},
  \bibinfo{author}{\bibfnamefont{M.}~\bibnamefont{Jaiswal}},
  \bibinfo{author}{\bibfnamefont{Z.}~\bibnamefont{Yi}},
  \bibinfo{author}{\bibfnamefont{S.~R.} \bibnamefont{Ali}},
  \bibinfo{author}{\bibfnamefont{G.}~\bibnamefont{G\"{u}ntherodt}},
  \bibinfo{author}{\bibfnamefont{B.~H.} \bibnamefont{Hong}},
  \bibnamefont{et~al.}, \bibinfo{journal}{Nano Lett.}
  \textbf{\bibinfo{volume}{11}}, \bibinfo{pages}{2363} (\bibinfo{year}{2011}).

\bibitem[{\citenamefont{Han and Kawakami}(2011)}]{PRL107_Han2011}
\bibinfo{author}{\bibfnamefont{W.}~\bibnamefont{Han}} \bibnamefont{and}
  \bibinfo{author}{\bibfnamefont{R.~K.} \bibnamefont{Kawakami}},
  \bibinfo{journal}{Phys. Rev. Lett.} \textbf{\bibinfo{volume}{107}},
  \bibinfo{pages}{047207} (\bibinfo{year}{2011}).

\bibitem[{\citenamefont{Yang et~al.}(2011)\citenamefont{Yang, Balakrishnan,
  Volmer, Avsar, Jaiswal, Samm, Ali, Pachoud, Zeng, Popinciuc
  et~al.}}]{PRL107_Yang2011}
\bibinfo{author}{\bibfnamefont{T.-Y.} \bibnamefont{Yang}},
  \bibinfo{author}{\bibfnamefont{J.}~\bibnamefont{Balakrishnan}},
  \bibinfo{author}{\bibfnamefont{F.}~\bibnamefont{Volmer}},
  \bibinfo{author}{\bibfnamefont{A.}~\bibnamefont{Avsar}},
  \bibinfo{author}{\bibfnamefont{M.}~\bibnamefont{Jaiswal}},
  \bibinfo{author}{\bibfnamefont{J.}~\bibnamefont{Samm}},
  \bibinfo{author}{\bibfnamefont{S.~R.} \bibnamefont{Ali}},
  \bibinfo{author}{\bibfnamefont{A.}~\bibnamefont{Pachoud}},
  \bibinfo{author}{\bibfnamefont{M.}~\bibnamefont{Zeng}},
  \bibinfo{author}{\bibfnamefont{M.}~\bibnamefont{Popinciuc}},
  \bibnamefont{et~al.}, \bibinfo{journal}{Phys. Rev. Lett.}
  \textbf{\bibinfo{volume}{107}}, \bibinfo{pages}{047206}
  (\bibinfo{year}{2011}).

\bibitem[{\citenamefont{Jo et~al.}(2011)\citenamefont{Jo, Ki, Jeong, Lee, and
  Kettemann}}]{PRB84_Jo2011}
\bibinfo{author}{\bibfnamefont{S.}~\bibnamefont{Jo}},
  \bibinfo{author}{\bibfnamefont{D.-K.} \bibnamefont{Ki}},
  \bibinfo{author}{\bibfnamefont{D.}~\bibnamefont{Jeong}},
  \bibinfo{author}{\bibfnamefont{H.-J.} \bibnamefont{Lee}}, \bibnamefont{and}
  \bibinfo{author}{\bibfnamefont{S.}~\bibnamefont{Kettemann}},
  \bibinfo{journal}{Phys. Rev. B} \textbf{\bibinfo{volume}{84}},
  \bibinfo{pages}{075453} (\bibinfo{year}{2011}).

\bibitem[{\citenamefont{Maassen et~al.}(2012)\citenamefont{Maassen, van~den
  Berg, IJbema, Fromm, Seyller, Yakimova, and van Wees}}]{NL12_Maassen2012}
\bibinfo{author}{\bibfnamefont{T.}~\bibnamefont{Maassen}},
  \bibinfo{author}{\bibfnamefont{J.~J.} \bibnamefont{van~den Berg}},
  \bibinfo{author}{\bibfnamefont{N.}~\bibnamefont{IJbema}},
  \bibinfo{author}{\bibfnamefont{F.}~\bibnamefont{Fromm}},
  \bibinfo{author}{\bibfnamefont{T.}~\bibnamefont{Seyller}},
  \bibinfo{author}{\bibfnamefont{R.}~\bibnamefont{Yakimova}}, \bibnamefont{and}
  \bibinfo{author}{\bibfnamefont{B.~J.} \bibnamefont{van Wees}},
  \bibinfo{journal}{Nano Lett.} \textbf{\bibinfo{volume}{12}},
  \bibinfo{pages}{1498} (\bibinfo{year}{2012}).

\bibitem[{\citenamefont{Guimarães et~al.}(2012)\citenamefont{Guimarães,
  Veligura, Zomer, Maassen, Vera-Marun, Tombros, and van
  Wees}}]{NL0_Guimaraes2012}
\bibinfo{author}{\bibfnamefont{M.~H.~D.} \bibnamefont{Guimarães}},
  \bibinfo{author}{\bibfnamefont{A.}~\bibnamefont{Veligura}},
  \bibinfo{author}{\bibfnamefont{P.~J.} \bibnamefont{Zomer}},
  \bibinfo{author}{\bibfnamefont{T.}~\bibnamefont{Maassen}},
  \bibinfo{author}{\bibfnamefont{I.~J.} \bibnamefont{Vera-Marun}},
  \bibinfo{author}{\bibfnamefont{N.}~\bibnamefont{Tombros}}, \bibnamefont{and}
  \bibinfo{author}{\bibfnamefont{B.~J.} \bibnamefont{van Wees}},
  \bibinfo{journal}{Nano Lett.} \textbf{\bibinfo{volume}{12}},
  \bibinfo{pages}{3512} (\bibinfo{year}{2012}).

\bibitem[{\citenamefont{Abel et~al.}(2012)\citenamefont{Abel, Matsubayashi,
  Murray, Dimitrakopoulos, Farmer, Afzali, Grill, Sung, and
  LaBella}}]{JoVSTB30_Abel2012}
\bibinfo{author}{\bibfnamefont{J.}~\bibnamefont{Abel}},
  \bibinfo{author}{\bibfnamefont{A.}~\bibnamefont{Matsubayashi}},
  \bibinfo{author}{\bibfnamefont{T.}~\bibnamefont{Murray}},
  \bibinfo{author}{\bibfnamefont{C.}~\bibnamefont{Dimitrakopoulos}},
  \bibinfo{author}{\bibfnamefont{D.~B.} \bibnamefont{Farmer}},
  \bibinfo{author}{\bibfnamefont{A.}~\bibnamefont{Afzali}},
  \bibinfo{author}{\bibfnamefont{A.}~\bibnamefont{Grill}},
  \bibinfo{author}{\bibfnamefont{C.~Y.} \bibnamefont{Sung}}, \bibnamefont{and}
  \bibinfo{author}{\bibfnamefont{V.~P.} \bibnamefont{LaBella}},
  \bibinfo{journal}{J. Vac. Sci. Technol. B} \textbf{\bibinfo{volume}{30}},
  \bibinfo{pages}{04E109} (\bibinfo{year}{2012}).

\bibitem[{\citenamefont{McCreary et~al.}(2012)\citenamefont{McCreary, Swartz,
  Han, Fabian, and Kawakami}}]{c_McCreary2012}
\bibinfo{author}{\bibfnamefont{K.~M.} \bibnamefont{McCreary}},
  \bibinfo{author}{\bibfnamefont{A.~G.} \bibnamefont{Swartz}},
  \bibinfo{author}{\bibfnamefont{W.}~\bibnamefont{Han}},
  \bibinfo{author}{\bibfnamefont{J.}~\bibnamefont{Fabian}}, \bibnamefont{and}
  \bibinfo{author}{\bibfnamefont{R.~K.} \bibnamefont{Kawakami}},
  \bibinfo{journal}{arXiv:1206.2628v1}  (\bibinfo{year}{2012}).

\bibitem[{\citenamefont{Wojtaszek et~al.}()\citenamefont{Wojtaszek, Vera-Marun,
  Maassen, and van Wees}}]{unpublished_Wojtaszek}
\bibinfo{author}{\bibfnamefont{M.}~\bibnamefont{Wojtaszek}},
  \bibinfo{author}{\bibfnamefont{I.~J.} \bibnamefont{Vera-Marun}},
  \bibinfo{author}{\bibfnamefont{T.}~\bibnamefont{Maassen}}, \bibnamefont{and}
  \bibinfo{author}{\bibfnamefont{B.~J.} \bibnamefont{van Wees}},
  \bibinfo{note}{submitted}.

\bibitem[{\citenamefont{Dlubak et~al.}(2012)\citenamefont{Dlubak, Martin,
  Deranlot, Servet, Xavier, Mattana, Sprinkle, Berger, De~Heer, Petroff
  et~al.}}]{NPaoP_Dlubak2012}
\bibinfo{author}{\bibfnamefont{B.}~\bibnamefont{Dlubak}},
  \bibinfo{author}{\bibfnamefont{M.-B.} \bibnamefont{Martin}},
  \bibinfo{author}{\bibfnamefont{C.}~\bibnamefont{Deranlot}},
  \bibinfo{author}{\bibfnamefont{B.}~\bibnamefont{Servet}},
  \bibinfo{author}{\bibfnamefont{S.}~\bibnamefont{Xavier}},
  \bibinfo{author}{\bibfnamefont{R.}~\bibnamefont{Mattana}},
  \bibinfo{author}{\bibfnamefont{M.}~\bibnamefont{Sprinkle}},
  \bibinfo{author}{\bibfnamefont{C.}~\bibnamefont{Berger}},
  \bibinfo{author}{\bibfnamefont{W.~A.} \bibnamefont{De~Heer}},
  \bibinfo{author}{\bibfnamefont{F.}~\bibnamefont{Petroff}},
  \bibnamefont{et~al.}, \bibinfo{journal}{Nature Phys.}
  \textbf{\bibinfo{volume}{8}}, \bibinfo{pages}{557} (\bibinfo{year}{2012}).

\bibitem[{\citenamefont{Virojanadara et~al.}(2008)\citenamefont{Virojanadara,
  Syv\"ajarvi, Yakimova, Johansson, Zakharov, and
  Balasubramanian}}]{PRB78_Virojanadara2008}
\bibinfo{author}{\bibfnamefont{C.}~\bibnamefont{Virojanadara}},
  \bibinfo{author}{\bibfnamefont{M.}~\bibnamefont{Syv\"ajarvi}},
  \bibinfo{author}{\bibfnamefont{R.}~\bibnamefont{Yakimova}},
  \bibinfo{author}{\bibfnamefont{L.~I.} \bibnamefont{Johansson}},
  \bibinfo{author}{\bibfnamefont{A.~A.} \bibnamefont{Zakharov}},
  \bibnamefont{and}
  \bibinfo{author}{\bibfnamefont{T.}~\bibnamefont{Balasubramanian}},
  \bibinfo{journal}{Phys. Rev. B} \textbf{\bibinfo{volume}{78}},
  \bibinfo{pages}{245403} (\bibinfo{year}{2008}).

\bibitem[{\citenamefont{Emtsev et~al.}(2009)\citenamefont{Emtsev, Bostwick,
  Horn, Jobst, Kellogg, Ley, McChesney, Ohta, Reshanov, Rohrl
  et~al.}}]{NM8_Emtsev2009}
\bibinfo{author}{\bibfnamefont{K.~V.} \bibnamefont{Emtsev}},
  \bibinfo{author}{\bibfnamefont{A.}~\bibnamefont{Bostwick}},
  \bibinfo{author}{\bibfnamefont{K.}~\bibnamefont{Horn}},
  \bibinfo{author}{\bibfnamefont{J.}~\bibnamefont{Jobst}},
  \bibinfo{author}{\bibfnamefont{G.~L.} \bibnamefont{Kellogg}},
  \bibinfo{author}{\bibfnamefont{L.}~\bibnamefont{Ley}},
  \bibinfo{author}{\bibfnamefont{J.~L.} \bibnamefont{McChesney}},
  \bibinfo{author}{\bibfnamefont{T.}~\bibnamefont{Ohta}},
  \bibinfo{author}{\bibfnamefont{S.~A.} \bibnamefont{Reshanov}},
  \bibinfo{author}{\bibfnamefont{J.}~\bibnamefont{Rohrl}},
  \bibnamefont{et~al.}, \bibinfo{journal}{Nature Mater.}
  \textbf{\bibinfo{volume}{8}}, \bibinfo{pages}{203} (\bibinfo{year}{2009}).

\bibitem[{foo({\natexlab{b}})}]{foot_reproduced}
\bibinfo{note}{The results for MLEG on SiC(0001) have been reproduced in our
  lab for 6H-SiC after obtaining the data on 4H-SiC published in Ref.~10}.

\bibitem[{\citenamefont{van~den Berg}()}]{unpublished_vdBerg}
\bibinfo{author}{\bibfnamefont{J.~J.} \bibnamefont{van~den Berg}},
  \bibinfo{note}{et al. in preparation}.

\bibitem[{\citenamefont{Riedl et~al.}(2009)\citenamefont{Riedl, Coletti,
  Iwasaki, Zakharov, and Starke}}]{PRL103_Riedl2009}
\bibinfo{author}{\bibfnamefont{C.}~\bibnamefont{Riedl}},
  \bibinfo{author}{\bibfnamefont{C.}~\bibnamefont{Coletti}},
  \bibinfo{author}{\bibfnamefont{T.}~\bibnamefont{Iwasaki}},
  \bibinfo{author}{\bibfnamefont{A.~A.} \bibnamefont{Zakharov}},
  \bibnamefont{and} \bibinfo{author}{\bibfnamefont{U.}~\bibnamefont{Starke}},
  \bibinfo{journal}{Phys. Rev. Lett.} \textbf{\bibinfo{volume}{103}},
  \bibinfo{pages}{246804} (\bibinfo{year}{2009}).

\bibitem[{\citenamefont{Speck et~al.}(2011)\citenamefont{Speck, Jobst, Fromm,
  Ostler, Waldmann, Hundhausen, Weber, and Seyller}}]{APL99_Speck2011}
\bibinfo{author}{\bibfnamefont{F.}~\bibnamefont{Speck}},
  \bibinfo{author}{\bibfnamefont{J.}~\bibnamefont{Jobst}},
  \bibinfo{author}{\bibfnamefont{F.}~\bibnamefont{Fromm}},
  \bibinfo{author}{\bibfnamefont{M.}~\bibnamefont{Ostler}},
  \bibinfo{author}{\bibfnamefont{D.}~\bibnamefont{Waldmann}},
  \bibinfo{author}{\bibfnamefont{M.}~\bibnamefont{Hundhausen}},
  \bibinfo{author}{\bibfnamefont{H.~B.} \bibnamefont{Weber}}, \bibnamefont{and}
  \bibinfo{author}{\bibfnamefont{T.}~\bibnamefont{Seyller}},
  \bibinfo{journal}{Appl. Phys. Lett.} \textbf{\bibinfo{volume}{99}},
  \bibinfo{eid}{122106} (\bibinfo{year}{2011}).

\bibitem[{foo({\natexlab{c}})}]{foot_tilt}
\bibinfo{note}{The spins are injected in the plane for small magnetic fields
  that do not tilt the magnetization of the injecting electrode out of plane.}

\bibitem[{foo({\natexlab{d}})}]{foot_noSpinflip}
\bibinfo{note}{We assume that the spin orientation is conserved in the coupling
  process described by this term.}

\bibitem[{foo({\natexlab{e}})}]{foot_SM}
\bibinfo{note}{See Supplemental Material at [URL will be inserted by publisher]
  for the discussion of the coupling rate between the localized states and the
  graphene channel.}

\bibitem[{foo({\natexlab{f}})}]{foot_length}
\bibinfo{note}{The distance between injector and detector in the two
  measurements is slightly different ($L=1.2~\mathrm{\mu m}$ vs.
  $L=1.5~\mathrm{\mu m}$) which has a minor influences on the width of the
  curve ($\mathrm{FWHM} \sim 1/L$). The difference in the amplitude of the
  measurements is not related to $\lambda_S$, which is comparable, but to a
  different polarization of the contacts. The data on MLEG shows no evidence
  that the change in the width from $W=0.7~\mathrm{\mu m}$ to $W=1~\mathrm{\mu
  m}$ has a strong influence on the measurements.}

\bibitem[{foo({\natexlab{g}})}]{foot_DDc}
\bibinfo{note}{The difference between $D$ and $D_C$ (and $D_C < D$) can be
  explained by the low contact resistances of the measurements $R_{C}\geq
  1.5~\mathrm{k\Omega}$ and the influence on the spin transport measurements as
  described in the publication on contact induced spin relaxation by T.~Maassen
  et al. (in preparation).}

\bibitem[{\citenamefont{Kopylov et~al.}(2010)\citenamefont{Kopylov, Tzalenchuk,
  Kubatkin, and Fal'ko}}]{APL97_Kopylov2010}
\bibinfo{author}{\bibfnamefont{S.}~\bibnamefont{Kopylov}},
  \bibinfo{author}{\bibfnamefont{A.}~\bibnamefont{Tzalenchuk}},
  \bibinfo{author}{\bibfnamefont{S.}~\bibnamefont{Kubatkin}}, \bibnamefont{and}
  \bibinfo{author}{\bibfnamefont{V.~I.} \bibnamefont{Fal'ko}},
  \bibinfo{journal}{Appl. Phys. Lett.} \textbf{\bibinfo{volume}{97}},
  \bibinfo{eid}{112109} (\bibinfo{year}{2010}).

\bibitem[{\citenamefont{Matsubara et~al.}(1990)\citenamefont{Matsubara,
  Sugihara, and Tsuzuku}}]{PRB41_Matsubara1990}
\bibinfo{author}{\bibfnamefont{K.}~\bibnamefont{Matsubara}},
  \bibinfo{author}{\bibfnamefont{K.}~\bibnamefont{Sugihara}}, \bibnamefont{and}
  \bibinfo{author}{\bibfnamefont{T.}~\bibnamefont{Tsuzuku}},
  \bibinfo{journal}{Phys. Rev. B} \textbf{\bibinfo{volume}{41}},
  \bibinfo{pages}{969} (\bibinfo{year}{1990}).

\end{thebibliography}
%

\pagebreak

\section{Supplemental Material for ``Localized states influence spin transport in epitaxial graphene''}
 
\textbf{Coupling rate between the localized states and the graphene channel}\\
 
To estimate the coupling rate $\Gamma$ between the localized states and the graphene channel, we can set up a simplified model based on the coupling between adjacent graphene layers in graphite, as these layers have the same or similar physical distance as the buffer layer to the graphene layer. \\
In graphite, the conductance in z-direction perpendicular to the layers is per layer $\sigma_{IL}=\sigma_{gr}/(\zeta d)$ where $\sigma_{gr}$ is the in-plane conductivity of a graphene layer, $d$ the distance between two layers (or between the graphene layer and the localized states) and $\zeta \approx 100$ the ratio between the conductivity within the layers and perpendicular to them \cite{PRB41_Matsubara1990}. \\
We can now calculate for a current $I_{IL}$ in z-direction:
\begin{equation}
I_{IL}=\frac{V \sigma_{IL} A}{d}=\frac{dQ}{dt}=e\frac{dN}{dt}=e \nu_{LS} \frac{d\mu}{dt} A
\label{eq:interlayer}
\end{equation}
Here $V=\mu/e$ is the voltage between the localized states and the channel, proportional to the difference in the chemical potential, $A$ the area through which the current flows, $Q$ is the total charge that flows, $N$ the number of charge carriers, $d$ is the distance to and $\nu_{LS}$ the density of states (DOS) of the localized states, and $e$ the electron charge.\\
Using the Einstein relation with $\nu_{gr}$ the DOS and $D$ the diffusion coefficient of the graphene channel we get:
\begin{equation}
V\frac{\nu_{gr}}{\nu_{LS}} \frac{D}{d^2} \frac{1}{\zeta} = \frac{dV}{dt}
\label{eq:interlayer4}
\end{equation}
This equation includes the ratio of the DOS of the localized states and the graphene channel $\eta=\nu_{LS}/\nu_{gr}$ that was discussed in the main text of this letter. \\
With the coupling rate $\Gamma \sim \frac{1}{V} \frac{dV}{dt}$ we receive :
\begin{equation}
\Gamma=\frac{1}{\eta} \frac{D}{d^2} \frac{1}{\zeta}.
\label{eq:interlayercoupling}
\end{equation}
With this model we get for bilayer graphene with $\zeta=100$, $\nu_{gr}=\nu_{LS}$, $d = 0.3~\mathrm{nm}$ and the typical graphene value $D\approx0.02~\mathrm{m^2/s}$
\begin{equation}
\Gamma_{BLG}\approx10^{15}~\mathrm{s^{-1}}.
\label{eq:GammaBLG}
\end{equation}
For our system we have $\eta \sim 50$ while the other parameters stay the same and get therefore 
\begin{equation}
\Gamma_{LS}\approx 2\times 10^{13}~\mathrm{s^{-1}}.
\label{eq:GammaLS}
\end{equation}
This value gives the order of magnitude of the coupling rate between the localized states and the graphene channel. With this value we are in the limit of strong coupling of the system as depicted in Fig.~2 of the main text.\\

\end{document}